\documentclass[pre,twocolumn,aps,showpacs]{revtex4}
\newcommand{\bec}[1]{\mbox{\boldmath $ #1$}}
\usepackage{graphicx}
\begin{document}
\title{Tangling clustering of inertial particles in stably stratified turbulence}
\medskip
\author{A. Eidelman}
\email{eidel@bgu.ac.il}
\author{T. Elperin}
\email{elperin@bgu.ac.il}
\homepage{http://www.bgu.ac.il/me/staff/tov}
\author{N. Kleeorin}
\email{nat@bgu.ac.il}
\author{B. Melnik}
\email{borism@bgu.ac.il}
\author{I. Rogachevskii}
\email{gary@bgu.ac.il}
\homepage{http://www.bgu.ac.il/~gary}

\medskip
\affiliation{The Pearlstone Center for Aeronautical Engineering
 Studies, Department of Mechanical Engineering,
 Ben-Gurion University of the Negev,
 P. O. Box 653, Beer-Sheva
84105, Israel}
\date{\today}
\begin{abstract}
We have predicted theoretically and detected in laboratory experiments a new type of particle clustering (namely, tangling clustering of inertial particles) in a stably stratified turbulence with imposed mean vertical temperature gradient. In the stratified turbulence a spatial distribution of the mean particle number density is nonuniform due to the phenomenon of turbulent thermal diffusion, i.e., the inertial particles are accumulated in the vicinity of the minimum of the mean temperature of the surrounding fluid, and a non-zero gradient of the mean particle number density, $\bec\nabla N$,  is formed. It causes  generation of fluctuations of the particle number density by tangling of the large-scale gradient, $\bec\nabla N$, by velocity fluctuations. In addition, the mean temperature gradient, $\bec\nabla T$, produces the temperature fluctuations by tangling of the large-scale gradient, $\bec\nabla T$, by velocity fluctuations. The anisotropic temperature fluctuations contribute to the two-point correlation function of the divergence of the particle velocity field, i.e., they increase the rate of formation of the particle clusters in small scales. We have demonstrated that in the laboratory stratified turbulence this tangling clustering is much more effective than a pure inertial clustering (preferential concentration) that has been observed in isothermal turbulence. In particular, in our experiments in oscillating grid isothermal turbulence in air without imposed mean temperature gradient, the inertial clustering is very weak for solid particles with the diameter $\approx 10 \, \mu$m and Reynolds numbers based on turbulent length scale and rms velocity, ${\rm Re} =250$. In the experiments the correlation function for the inertial clustering in isothermal turbulence is much smaller than that for the tangling clustering in non-isothermal turbulence.  The size of the tangling clusters is of the order of several Kolmogorov length scales. The clustering described in our study is found for inertial particles with small Stokes numbers and with the material density that is much larger than the fluid density. Our theoretical predictions are in a good agreement with the obtained experimental results.
\end{abstract}


\maketitle

\section{Introduction}

Laboratory experiments \cite{EF94,FK94,WA00}, numerical simulations \cite{WM93,MC96,SC97,S01} and atmospheric observations \cite{KM93,VY00,KS01,S03,BH03,KPE07,LS07} revealed small-scale long-living inhomogeneities (clusters) in spatial distribution of particles in different turbulent flows. The origin of these inhomogeneities is not always clear but their influence on particle transport and mixing can be hardly overestimated. In particular, turbulence causes formation of small-scale droplet inhomogeneities, increases the relative droplet velocity and affects the hydrodynamic droplet interactions. All these effects can enhance the rate of droplet collisions (see reviews \cite{VY00,S03,KPE07}).
It is known that atmospheric clouds are regions with strong turbulence, and the turbulence effects are of a great importance for understanding of rain formation in atmospheric clouds, e.g., these effects can cause the droplet spectrum broadening and acceleration of raindrop formation \cite{S03,VY00,KPE07,RC00}.

Different kinds of particle clustering, i.e., the preferential concentration of inertial particles have been studied in a number of numerical simulations \cite{HC01,B03,BL04,CK04,FP04,CC05,BE05,CG06,BC07,BB07,YG07,AC08,ACW08,GP09,BB10}, laboratory experiments (see review \cite{WA09} and \cite{AC02,WH05,AG06,DF06,SA08,XB08,SSA08}) and analytical investigations \cite{EKRC96,EKRC00,BF01,EKR02,ZA03,DM05,MW05,WM05,WM06,GDH05,EKR07,OV07,FM07,FH08,OL10}.

Inertial clustering in isothermal and nearly isotropic turbulence with the Reynolds numbers ${\rm Re} = u_0 \, \ell_0 / \nu \sim 10^3$ produced by fans at the corners of the box has been observed in laboratory experiments \cite{SA08} with hollow glass spheres with a mean diameter of $6 \, \mu$m in air. Here $u_0$ is the characteristic turbulent velocity in the maximum scale of turbulent motions $\ell_0$ and $\nu$ is the kinematic viscosity. The three-dimensional radial distribution function (RDF) as a statistical measure of clustering, has been determined from the particle position field using three-dimensional digital holographic particle imaging. These experiments reveled inertial particle clustering at the dissipation scales. The detailed comparisons between experiments and direct numerical simulations of inertial particle clustering in \cite{SA08} have demonstrated a good agreement.

The quantitative measurements of inertial clustering have been also performed in \cite{WH05,SSA08}. In particular, the two-dimensional RDF of particles suspended in a turbulence flow with the Reynolds numbers ${\rm Re} \sim 10^3$ has been measured in \cite{WH05} by shining a laser sheet at the particles (the glass particles with the sizes $20 \, \mu$m and $50 \, \mu$m, the lycopodium particles with the size $25 \, \mu$m), whereby particle locations have been determined by a CCD camera. The experiments have been conducted in the spherical turbulence chamber with eight synthetic jet actuators which create homogeneous and isotropic turbulence with no mean flow at the center of the chamber. The experiments conducted in \cite{WH05} have shown that the inertial clustering is most effective for particles with Stokes numbers near unity and the size of the clusters is of the order of ten Kolmogorov length scales.

A one-dimensional RDF has been measured in \cite{SSA08} by sampling droplet arrivals at a fixed volume in a wind tunnel, whereby the arrival statistics of water droplets with the mean size $22 \, \mu$m in turbulence with the Reynolds numbers ${\rm Re} \sim 10^4$ has been used to determine the one-dimensional RDF. In the experiments described in \cite{SSA08} a phase Doppler interferometer downstream of the active grid and a spray system have been used, and strong inertial clustering has been observed. Remarkably, the similar experimental findings have been reported in \cite{WH05,SA08} using completely different experimental set-ups.

The mechanism of inertial clustering of particles in isothermal turbulence is as follows \cite{M87}. The particles inside the turbulent eddies are carried out to the boundary regions between the eddies by the inertial forces. This mechanism of the inertial clustering  acts in all scales of turbulence, and is more pronounced in small scales.

The goal of this study is to investigate experimentally and theoretically a new type of particle clustering, namely tangling clustering of inertial particles in stably stratified turbulence with imposed mean vertical temperature gradient. In experimental study of particle clustering we use Particle Image Velocimetry to determine the turbulent velocity field, an image processing technique to determine the spatial distribution of particles, and a specially designed temperature probe with twelve sensitive thermocouples for measurements of the temperature field. The clustering described in our study is found for inertial particles with small Stokes numbers and with material density that is much larger than the fluid density.

In the stratified turbulence a spatial distribution of the mean particle number density is nonuniform due to phenomenon of turbulent thermal diffusion \cite{EKR96,EKR00}. In particular, the inertial particles are accumulated in the vicinity of the minimum of the mean temperature of the surrounding fluid, which causes formation a non-zero gradient of the mean particle number density, $\bec\nabla N$.
The phenomenon of turbulent thermal diffusion has been predicted theoretically in \cite{EKR96}, detected in the laboratory experiments in stably and unstably stratified turbulent flows in \cite{EEKR04,BEE04,EEKR06a,EEKR06b} and observed in atmospheric turbulence in \cite{SEKR09}.

Fluctuations of the particle number density can be generated by tangling of the gradient, $\bec\nabla N$, of the mean particle number density by velocity fluctuations \cite{EKR95}. On the other hand, the imposed mean temperature gradient, $\bec\nabla T$, results in generation of the anisotropic temperature fluctuations by tangling of this large-scale gradient, $\bec\nabla T$, by velocity fluctuations. These temperature fluctuations may contribute to the two-point correlation function of the divergence of the particle velocity field. The latter enhances the rate of formation of particle clusters in small scales by the tangling mechanism.

The tangling mechanism is universal and independent of the way of generation of turbulence for large Reynolds numbers. For instance, tangling of the gradient of the large-scale velocity shear produces anisotropic velocity fluctuations \cite{L67,WC72} which are responsible for different phenomena: formation of large-scale coherent structures in a turbulent convection \cite{EKRZ02}, excitation of the large-scale inertial waves in a rotating inhomogeneous turbulence \cite{EGKR05}, generation of large-scale vorticity \cite{EKR03} and large-scale magnetic field \cite{RK03} in a sheared turbulence.

The paper is organized as follows. Section II
describes the experimental set-up for a laboratory study of the tangling clustering of inertial particles in stably stratified turbulence. The data processing and experimental results are presented in Section III. The theoretical analysis and comparison with experimental results are performed in Section IV. Finally, conclusions are drawn in Section V.

\section{Experimental set-up}

\begin{figure}
\centering
\includegraphics[width=8cm]{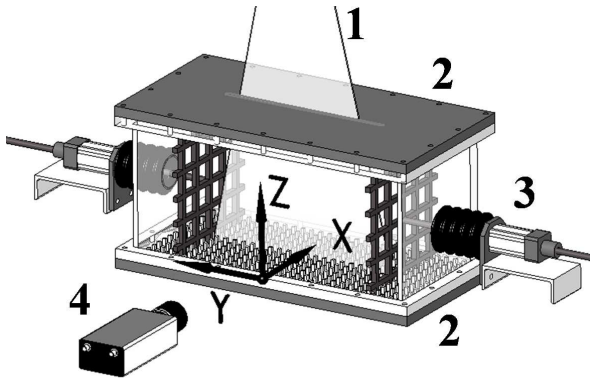}
\caption{\label{Fig1} Experimental set-up: (1) -laser light sheet; (2) - heat exchangers; (3) - grid driver; (4) - digital CCD camera.}
\end{figure}

The experiments were carried out in a turbulence generated by
oscillating grids in air. The test section of the oscillating grids
turbulence generator was constructed as a rectangular chamber with
dimensions $29 \times 58 \times 29$ cm$^3$ (see Fig.~\ref{Fig1}). Pairs of vertically oriented grids with bars arranged in a square array (with
a mesh size 5 cm) are attached to the right and left horizontal
rods driven by speed-controlled motors. The grids are positioned at a distance of two grid meshes from
the chamber walls parallel to them. Both grids are operated at the
same amplitude of $61$ mm, at a random phase and at the same
frequency varied in the range from $2.2$ Hz to $16.5$ Hz. Here we
use the following system of coordinates: $Z$ is the vertical axis,
the $Y$-axis is perpendicular to the grids and the $XZ$-plane is
parallel to the grids.

A mean temperature gradient in the turbulent flow was formed with
two aluminium heat exchangers attached to the bottom and top walls
of the chamber. We performed experiments in stably stratified fluid flow (the cold bottom and hot top walls of the chamber). In order to improve heat transfer in the boundary layers at the walls we used two heat exchangers with rectangular fins $(3 \times 3 \times 15$ mm$^3$, see Fig.~1) which affords a mean temperature gradient 118 K/m at a mean temperature of about 308 K. All experiments were conducted at the same temperature difference between the top and bottom walls $\Delta T = 50$ K (e.g., the bottom wall temperature was 283 K and the top wall temperature was 333 K).

The temperature field was measured with a temperature probe equipped
with twelve E-thermocouples (with the diameter of 0.13 mm and the
sensitivity of $65 \, \mu$V/K) attached to a vertical rod with a
diameter 4 mm. The spacing between thermocouples along the rod was
22 mm. Each thermocouple was inserted into a 1 mm diameter and 45 mm
long case. A tip of a thermocouple protruded at the length of 15 mm
out of the case. The mean temperature was measured for 5 rod
positions with 50 mm intervals in the horizontal direction, i.e., at
60 locations in a flow. The exact position of each thermocouple was
measured using images captured with the optical system employed in
PIV measurements. A sequence of 1024 temperature readings (each
reading was averaged over 15 instantaneous measurements) for every
thermocouple at every rod position was recorded and processed using
the developed software based on LabVIEW 7.0.

The turbulent velocity field was measured using a digital Particle
Image Velocimetry (PIV) system with LaVision Flow Master III (see,
e.g., \cite{AD91,RWK98,W00}). A double-pulsed Nd-YAG laser (Continuum
Surelite $ 2 \times 170$ mJ) is used for light sheet formation. Light sheet optics comprise spherical and cylindrical Galilei telescopes with tuneable divergence and adjustable focus length. We employed a progressive-scan 12 Bit
digital CCD camera (pixels with a size $6.7 \,\mu$m $\times 6.7 \, \mu$m each) with dual frame technique for cross-correlation processing of
captured images. The tracer used for PIV measurements was incense
smoke with sub-micron particles (with the material density $ \rho_{\rm tr} \approx 1 $ g/cm$^3$), which was produced by high temperature sublimation of solid incense particles. Analysis of smoke particles using a microscope (Nikon, Epiphot with an amplification 560) and PM-300 portable laser particulate analyzer showed that these particles have an
approximately spherical shape with a mean diameter of $0.7 \mu$m.

We determined the mean and the r.m.s.~velocities, two-point
correlation functions and an integral scale of turbulence from the
measured velocity fields. Series of 130 pairs of images, acquired
with a frequency of 4 Hz, were stored for calculating velocity maps
and for ensemble and spatial averaging of turbulence
characteristics. The center of the measurement region coincides with
the center of the chamber. We measured velocity for flow areas from
$50 \times 50$ mm$^2$ up to $212 \times 212$ mm$^2$ with a spatial
resolution of $1024 \times 1024$ pixels. This corresponds to a
spatial resolution from 48 $\mu$m / pixel up to 207 $\mu$m / pixel.
These measurement regions were analyzed with interrogation windows
of $32 \times 32$ or $16 \times 16$ pixels, respectively.

In every interrogation window a velocity vector was determined from which velocity maps comprising $32 \times 32$ or $64 \times 64$ vectors
were constructed. The mean and r.m.s. velocities for every point of
a velocity map (1024 points) were calculated by averaging over 130
independent maps, and then they were averaged over 1024 points. The
two-point correlation functions of the velocity field were
calculated for every point of the central part of the velocity map
(with $16 \times 16$ vectors) by averaging over 130 independent
velocity maps, and then they were averaged over 256 points. An
integral scale $\ell_0$ of turbulence was determined from the two-point
correlation functions of the velocity field.

Particle spatial distribution was determined using digital Particle
Image Velocimetry (PIV) system. In particular, the effect of Mie
light scattering by particles was used to determine the particle
spatial distribution in the flow (see, e.g., \cite{G01}).
In the experiments we probed the central $212 \times 212$ mm$^2$
region in the chamber. The mean intensity of scattered light was
determined in $32 \times 16$ interrogation windows with the size $32 \times 64$ pixels. The vertical distribution of the intensity of
the scattered light was determined in 16 vertical strips composed of
32 interrogation windows. The light radiation energy flux scattered
by small particles is $E_s \propto E_0 \Psi(\pi d_p/\lambda; a_0;
N)$, where $E_0 \propto \pi d_p^2 / 4$ is the energy flux incident
at the particle, $d_p$ is the particle diameter, $\lambda$ is the
wavelength, $a_0$ is the index of refraction and $\Psi$ is the
scattering function. For wavelengths $\lambda$ which are larger than
the particle  perimeter $(\lambda > \pi d_p)$, the function $\Psi$
is given by Rayleigh's law, $\Psi \propto d_p^4$. If the wavelength
is small, the function $\Psi$ tends to be independent of $d_p$ and
$\lambda$. In the general case the function $\Psi$ is given by Mie's
equations (see, e.g., \cite{BH83}, Chapter 4).
The scattered light energy flux incident on the CCD camera probe
(producing proportional charge in every CCD pixel) is proportional
to the particle number density $n$, i.e., $E_s \propto E_0 \, n \,
(\pi d_p^2 / 4)$.

In order to characterize the spatial distribution of particle number
density $n \propto E^T / E$ in the non-isothermal flow, the
distribution of the scattered light intensity $E$ measured in the
isothermal case was used for the normalization of the scattered
light intensity $E^T$ obtained in a non-isothermal flow under the
same conditions. The scattered light intensities $E^T$ and $E$ in
each experiment were normalized by corresponding scattered light
intensities averaged over the vertical coordinate. Mie scattering is
not affected by temperature change because it depends on the
electric permittivity of particles, the particle size and the laser
light wave length. The temperature effect on these characteristics
is negligibly small. Similar experimental set-up and data processing procedure were used in experimental study of different aspects of turbulent convection \cite{EEKR06c,BEKR09} and in \cite{EEKR04,BEE04,EEKR06a,EEKR06b} for investigating the phenomenon of turbulent thermal diffusion \cite{EKR96,EKR00}.

For experimental study of particle clustering we used  hollow
borosilicate glass particles having an approximately spherical shape, a mean diameter of $10 \, \mu$m and the material density $ \rho_p \approx 1.4 $ g/cm$^3$. These particles have been injected in the chamber using an air jet in order to improve particle mixing and prevent from particle agglomeration.

\section{Data processing and experimental results}

All experiments for study of the particle clustering have been performed at the frequency of grid oscillations $f=10.4$ Hz.
The turbulent flow parameters in the oscillating grids turbulence generator at $f=10.4$ Hz are as follows: the r.m.s. velocity is $u_0 =\sqrt{\langle {\bf u}^2 \rangle} = 12$ cm/s, the integral (maximum) scale of turbulence is $\ell_0 = 3.2$ cm, the Reynolds numbers ${\rm Re} = u_0 \, \ell_0 / \nu =250$, the Kolmogorov length scale is $\ell_\eta = \ell_0 / {\rm Re}^{3/4} = 510 \, \mu$m and the Kolmogorov time scale
$\tau_\eta =\tau_0 / {\rm Re}^{1/2} = 1.7 \times 10^{-2}$ s, where $\tau_0 = \ell_0/u_0$.  The Stokes time for the particles with the diameter $d_p = 10 \, \mu$m is $\tau_s = 10^{-3}$ s, the Stokes number ${\rm St} =\tau_s/\tau_\eta = 5.9 \times 10^{-2}$, the coefficient of molecular diffusion $D_m = 1.4 \times 10^{-8}$ cm$^2$ /s and the Peclet number Pe$=u_0 \, \ell_0 / D_m =3 \times 10^9$.

The velocity measurements have shown that there is a slight difference in the velocity components (about 9\%) caused by an anisotropy of forcing of the grid oscillating turbulence. The inhomogeneity of turbulence in the core of fluid flow is weak. We have found a weak mean flow in the form of two large toroidal structures parallel and adjacent to the grids. The interaction of these structures results in a symmetric mean flow that is sensitive to the parameters of grids adjustment. We studied the parameters that affect the mean flow, e.g., the grids distance to the walls of the chamber, the angles of the grids planes with the axes of their oscillation. This study allowed us to expand the
central region with homogeneous turbulence by inserting partitions
behind the grids. The measured r.m.s. velocity was 5 times higher
than the characteristic mean velocity in the core of the flow.

For analysis of particle clustering we use the radial distribution function (RDF), $G({\bf R}) =\langle n(t,{\bf x}) n(t,{\bf y}) \rangle / N(t,{\bf x}) N(t,{\bf y})$, that is the conditional probability density of finding a second particle at a given separation distance from a test particle (see, e.g., \cite{LL80}), where  $n(t,{\bf x})$ is the instantaneous number density of particles,  $N(t,{\bf x}) = \langle n(t,{\bf x})\rangle$ is the mean number density of particles, the angular brackets denote ensemble averaging and ${\bf R}={\bf y}-{\bf x}$. The RDF can be estimated from a field of $M$ particles by binning the particle pairs according to their separation distance, so that the function $G({\bf R})$ is estimated as follows:
\begin{eqnarray}
G({\bf R}) \approx {N_{\Delta V}^{(p)} / \Delta V \over N_V^{(p)} / V} \;
\label{E1}
\end{eqnarray}
(see, e.g., \cite{HOO91}), where $N_{\Delta V}^{(p)}$ is the number of particle pairs separated by a distance $R \pm \Delta R/2$, $\; \Delta V$ is the volume of the spherical shell located between $R \pm \Delta R/2$, $\; N_V^{(p)}= M (M-1)/2$ is the total number of pairs and $V$ is the total volume of the probed region.

In our experiments we employ the PIV system in order to determine the particle spatial distribution. Since we use the two-dimensional images, Eq.~(\ref{E1}) is modified as follows:
\begin{eqnarray}
G({\bf R}) \approx {N_{\Delta S}^{(p)} / \Delta S \over N_S^{(p)} / S} \;,
\label{E2}
\end{eqnarray}
where $\Delta S = \pi [(R + \Delta R/2)^2 - (R - \Delta R/2)^2]$ is the area of the annular domain located between $R \pm \Delta R/2$, $\, S$ is the area of the part of the image with the radius $R_{\rm max}$ that was used in data processing in order to exclude the edge effects.
In our experiments with the image sizes $20 \times 20$ cm$^2$ and $5 \times 5$ cm$^2$, the maximum radius $R_{\rm max} = 0.8$ cm and $R_{\rm max} = 0.2$ cm, respectively. The total number of particles in the image $5 \times 5$ cm$^2$ is of the order of $M \sim 2 \times 10^4$.
Using the thickness of the laser light sheet $d = 0.2$ cm we can estimate the effective 3D particle mean number density in the experiments as $N \sim 4 \times 10^3$ cm$^{-3}$.

The measured radial distribution function has been used for determining the two-point correlation function of the particle number density, $\Phi(t,{\bf R})=\langle \Theta(t,{\bf x}) \Theta(t,{\bf x}+{\bf R}) \rangle$, where $\Theta(t,{\bf x}) = n(t,{\bf x}) - N(t,{\bf x})$ is the deviation of the instantaneous number density of particles $n(t,{\bf x})$ from the mean number density of particles $N(t,{\bf x})$. The two-point correlation function of the particle number density
is given by
\begin{eqnarray}
\Phi(t,{\bf R}) = N^2 \, [G(t,{\bf R}) - 1] \;,
\label{EE2}
\end{eqnarray}
(see, e.g., \cite{LL80}). Note that in our experiments the correlation function has been normalized by squared mean number density of particles in every image. We perform the double averaging over all particles in the image and then over ensemble of 50 images, which allow us to increase the accuracy of determining the two-point correlation function of the particle number density. Therefore, particle clustering in our study is understood in the statistical sense by applying an ensemble averaging over many images with the instantaneous particle distributions.

Since a typical size of a particle cluster is of the order of several Kolmogorov length-scales of turbulence (see, e.g., \cite{EKRC96,EKR02,SA08}), we have to use a sub-pixel resolution in the data analysis. In our experiments one pixel is of the order of $1/3$ of the Kolmogorov length scale of turbulence. On the other hand, the size of the analyzed region in the image cannot be reduced strongly, because a number of particles in the analyzed region of the image should be large in order to provide a good statistics.

In order to attain the sub-pixel resolution we employ the following method: (i) we determine the response function for the CCD camera used in the PIV system by analyzing the light intensity distribution in the image for single particles located at the center of the pixel in the form of the Gaussian distribution with $\sigma_\ast = 0.81$ pixels; (ii) segmentation of the image using a threshold technique; (iii) identification of particle locations in the segments by least-square fitting of the recorded light intensity distribution and the light intensity distribution caused by superposition of the Gaussian distributions at the particle locations. This procedure has been tested using artificial computer-generated images. The error in the determining the particle coordinates is of the order of $3 \%$.

\begin{figure}
\vspace*{2mm} \centering
\includegraphics[width=9cm]{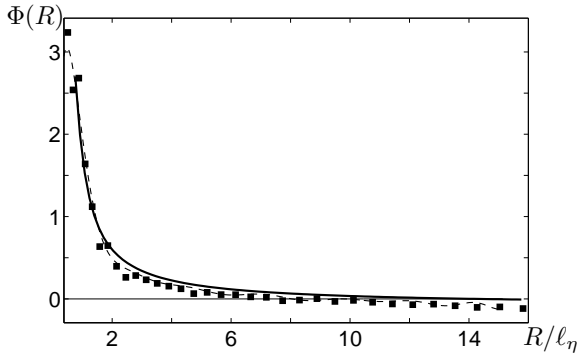}
\caption{\label{Fig2} Normalized two-point second-order correlation function $\Phi(R)$ determined in our experiments (filled squares) and the least-square fit for the experimental results (dashed line). Normalized two-point second-order correlation function $\Phi(R)$ determined from our model for $\sigma_{_{T}}=1$, $\; a_c=1$ (solid line).}
\end{figure}

This approach allows us to determine in the experiments the two-point second-order correlation function $\Phi(R)$ of the particle number density. For instance, in Fig.~\ref{Fig2} we plot the normalized two-point second-order correlation function $\Phi(R)$ determined in our experiments, which are performed in the air flow with imposed mean temperature gradient, i.e., for the stably stratified turbulence with the temperature difference,  $\Delta T = 50$ K, between the top and bottom walls of the chamber.

\begin{figure}
\vspace*{2mm} \centering
\includegraphics[width=8.5cm]{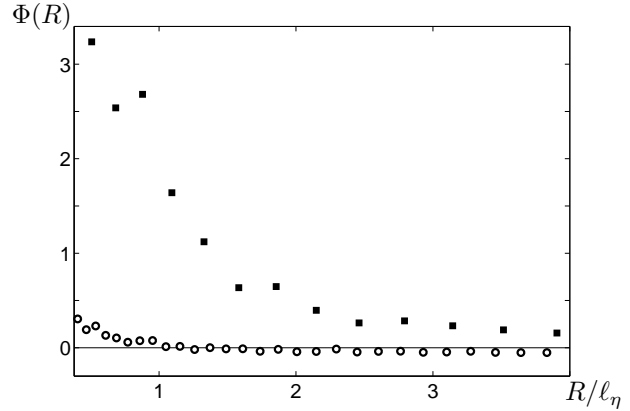}
\caption{\label{Fig3} Normalized two-point second-order correlation function $\Phi(R)$ determined in our experiments: (i) for isothermal turbulence (unfilled circles); (ii) for non-isothermal turbulence (filled squares as in  Fig.~\ref{Fig2}).}
\end{figure}

On the other hand, we have found that in the experiments in isothermal turbulence without imposed mean temperature gradient, the inertial particle clustering is very weak, i.e., the correlation function $\Phi(R)$ for the inertial clustering is much smaller than that for the tangling clustering. Indeed, in Fig.~\ref{Fig3} we compare the correlation function $\Phi(R)$ for the isothermal turbulence (unfilled circles) with that for the turbulence with imposed mean temperature gradient (filled squares). For instance, for the isothermal turbulence $\Phi_{\rm in}(R=0.4 \ell_\eta) = 0.3$, while for the turbulence with imposed mean temperature gradient $\Phi_{\rm tan}(R=0.4 \ell_\eta) = 3.3$ (the tangling clustering). The minimum distance $R_0$ at which the correlation function $\Phi(R)$ approaches zero is $R_0=\ell_\eta$ for the isothermal turbulence and $R_0=7.6 \ell_\eta$ for the tangling clustering.

In the next section we perform theoretical study of this new kind of particle tangling clustering and compare theoretical predictions with the experimental results.

\section{Theoretical analysis and comparison with experimental results}

Let us consider the two-point second-order correlation function of the particle number density fluctuations generated by tangling of the gradient of the mean particle number density by the turbulent velocity field. This gradient is formed due to phenomenon of turbulent thermal diffusion in stably stratified turbulence.

\subsection{General consideration}

The equation for the number density $n(t,{\bf x})$ of particles advected by a fluid velocity field reads:
\begin{eqnarray}
\frac{\partial n}{\partial t} + {\bf \nabla \cdot} (n \,{\bf v}) =
D_m\, \Delta n \;,
 \label{B1}
\end{eqnarray}
where $D_m$ is the coefficient of molecular (Brownian) diffusion and ${\bf v}(t,{\bf x})$ is the particle velocity field. Equation~(\ref{B1}) implies conservation of the total number of particles in a closed volume. The equation for fluctuations of the particle number density, $\Theta(t,{\bf x})= n(t,{\bf x}) - N(t,{\bf x})$, reads:
\begin{eqnarray}
\frac{\partial \Theta}{\partial t} &+& {\bf \nabla \cdot} (\Theta \,{\bf v}- \langle \Theta \,{\bf v}\rangle) = D_m\, \Delta \Theta
\nonumber\\
&& - ({\bf v} {\bf \cdot \nabla}) N - N \, {\bf \nabla \cdot} \,{\bf v} \;,
 \label{RB1}
\end{eqnarray}
(see, e.g., \cite{BA71,BA59}). Using Eq.~(\ref{RB1}) we derive equation for the evolution of the two-point second-order correlation function of the particle number density, $\Phi(t,{\bf R})=\langle \Theta(t,{\bf x}) \Theta(t,{\bf x}+{\bf R}) \rangle$:
\begin{eqnarray}
{\partial \Phi \over \partial t} &=& \big[B({\bf R}) + 2 {\bf U}^{(A)}({\bf R})\cdot \bec{\nabla} + \hat D_{ij}({\bf R}) \nabla_{i}
\nabla_{j}\big] \, \Phi(t,{\bf R})
\nonumber\\
&&+ I({\bf R}) \,,
\label{B2}
\end{eqnarray}
(see \cite{EKR02}), where ${\bf U}^{(A)}({\bf R}) = (1/2) \, \big[{\bf U}({\bf R}) - {\bf U}(-{\bf R})\big]$,
\begin{eqnarray}
\hat D_{ij}&=& 2 D_m \delta_{ij} + D_{ij}^{^{T}}(0) - D_{ij}^{^{T}}({\bf R}) ,
 \label{AL1}\\
D_{ij}^{^{T}}({\bf R}) &=& 2 \int_{0}^{\infty} \langle v_{i}\big[0,\bec{\xi}(t,{\bf x}|0)\big] \, v_{j}\big[\tau,\bec{\xi}(t,{\bf x}+{\bf R}|\tau)\big] \rangle \,d \tau ,
\nonumber\\
 \label{AL2}\\
B({\bf R}) &=& 2 \int_{0}^{\infty} \langle b\big[0,\bec{\xi}(t,{\bf x}|0)\big] \,b\big[\tau,\bec{\xi}(t,{\bf x}+{\bf R}|\tau)\big] \rangle \,d \tau ,
\nonumber\\
 \label{AL3}\\
U_{i}({\bf R}) &=& -2 \int_{0}^{\infty} \langle v_{i}\big[0,\bec{\xi}(t,{\bf x}|0)\big] \,b\big[\tau,\bec{\xi}(t,{\bf x}+{\bf R}|\tau)\big] \rangle \,d \tau .
\nonumber\\
 \label{AL4}
\end{eqnarray}
$b={\rm div} \, {\bf v}$,  $\, D_{ij}^{^{T}}({\bf R})$ is the scale-dependent turbulent diffusion tensor, $\delta_{ij}$ is the Kronecker tensor, $I({\bf R})$ is the source of particle number density fluctuations and $\langle ... \rangle$ denotes averaging over the statistics of turbulent velocity field and the Wiener process ${\bf w}(t)$. The Wiener trajectory $\bec{\xi}(t,{\bf x}|s)$ (which is often called the Wiener path) in the expressions for the turbulent diffusion tensor $D_{ij}^{^{T}} ({\bf R})$ and other transport coefficients is defined as follows:
\begin{eqnarray}
\bec{\xi}(t,{\bf x}|s) &=&{\bf x} - \int^{t}_{s} {\bf v}
[\tau,\bec{\xi}(t,{\bf x}|\tau)] \,\,d \tau - \sqrt{2 D_m} \, {\bf
w}(t-s) ,
\nonumber\\
 \label{B33}
\end{eqnarray}
where ${\bf w}(t)$ is the Wiener random process which describes the
Brownian motion (molecular diffusion). The Wiener random process
${\bf w}(t)$ is defined by the following properties: $\langle {\bf
w}(t) \rangle_{\bf w}=0\,, $ $\, \langle w_i(t+\tau) w_j(t)
\rangle_{\bf w}= \tau \delta _{ij}$, and $ \langle \dots
\rangle_{\bf w} $ denotes the mathematical expectation over the
statistics of the Wiener process. The velocity $v_i[\tau,
\bec{\xi}(t,{\bf x}|\tau)]$ describes the Eulerian velocity
calculated at the Wiener trajectory.

The source function $I({\bf R})$ in Eq.~(\ref{B2}) is related to the last two terms $- ({\bf v} {\bf \cdot \nabla}) N - N \, {\bf \nabla \cdot} \,{\bf v}$ in the right hand side of Eq.~(\ref{RB1}). In particular, when $\bec {\nabla} N \not= 0$ the nonzero source $I$ results in generation of fluctuations of the particle number density caused by tangling of the gradient of the mean particle number density by the turbulent velocity field. The source function $I({\bf R})$ is given by:
\begin{eqnarray}
I({\bf R}) &=& B({\bf R}) N^2 + {\bf U}^{(S)}({\bf R}) \cdot \bec{\nabla} N^2
\nonumber\\
&&+ {3 \over 4} D_{ij}^{^{T}}({\bf R}) \, (\nabla_{i} N)
\, (\nabla_{j}N) ,
\label{IB1}
\end{eqnarray}
where ${\bf U}^{(S)}({\bf R}) = (1/2) \, \big[{\bf U}({\bf R}) + {\bf U}(-{\bf R})\big]$ and we have taken into account that $\nabla_i^{({\bf x})} \nabla_j^{({\bf y})} N(t,{\bf x}) N(t,{\bf y}) = (3/4) \, (\nabla_{i} N) \, (\nabla_{j}N)$ and $\nabla_i \equiv \nabla_i^{({\bf R})}$.

The meaning of the turbulent transport coefficients $B({\bf R})$ and ${\bf U}({\bf R})$ is as follows. The function $B({\bf R})$ is determined by the compressibility of the particle velocity field. The vector ${\bf U}({\bf R})$ determines a scale-dependent drift velocity which describes transport of fluctuations of particle number density from smaller scales to larger scales, i.e., in the regions with larger turbulent diffusion. The scale-dependent tensor of turbulent diffusion $D_{ij}^{^{T}}({\bf R})$ in very small scales is equal to
the tensor of the molecular (Brownian) diffusion, while in the
vicinity of the maximum scale of turbulent motions this tensor
coincides with the regular tensor of turbulent diffusion.

If the turbulent velocity field is not delta-correlated in time (e.g., the correlation time is small yet finite), the tensor of turbulent diffusion, $D_{ij}^{^{T}} ({\bf R})$, is compressible, i.e.,
$(\partial / \partial R_i) D_{ij}^{^{T}} ({\bf R}) \not= 0$ (for details see \cite{EKR02}). The parameter $\sigma_{_{T}}$ that characterizes the degree of compressibility of the tensor of turbulent diffusion is defined as follows:
\begin{eqnarray}
\sigma_{_{T}} \equiv \frac{\nabla_i \nabla_j D^{^{\rm T}}_{ij}({\bf R})}{\nabla_i\nabla_j D^{^{\rm T}} _{mn}({\bf R}) \epsilon_{imp}\epsilon_{jnp} } \approx {\langle (\bec{\nabla} {\bf \cdot} \bec{\tilde \xi})^{2} \rangle \over \langle (\bec{\nabla} {\bf \times} \bec{\tilde \xi})^{2} \rangle} \;,
 \label{S1}
\end{eqnarray}
where $\epsilon_{ijk}$ is the fully antisymmetric Levi-Civita unit tensor, $\bec{\tilde \xi} =\bec{\xi} - {\bf x}$ with $|t-s| \gg \tau_c$. Here $\tau_c$ is the characteristic turbulent time,
\begin{eqnarray}
\bec{\nabla} {\bf \cdot} \bec{\tilde \xi} = -\int^{t}_{s} {\partial u_i
[\tau,\bec{\xi}(t,{\bf x}|\tau)] \over \partial \xi_m} \, {\partial \xi_m \over \partial x_i} \,\,d \tau
\;,
 \label{AD7}
 \end{eqnarray}
and
\begin{eqnarray}
{\partial \xi_m \over \partial x_i} =\delta_{mi} -\int^{t}_{\tau} {\partial u_m [\tau',\bec{\xi}(t,{\bf x}|\tau')] \over \partial \xi_n} \, {\partial \xi_n \over \partial x_i} \,\,d \tau'
\; .
 \label{AD8}
 \end{eqnarray}
When the turbulent velocity field is a delta-correlated in time random process, $\bec{\nabla} {\bf \cdot} \bec{\tilde \xi} = - (\bec{\nabla} {\bf \cdot} {\bf u}) \, (t-s)$ and $\bec{\nabla} {\bf \times} \bec{\tilde \xi} = - (\bec{\nabla} {\bf \times} {\bf u}) \, (t-s)$. Therefore, in this case the parameter $\sigma_{_{T}}$ is given by
\begin{eqnarray}
\sigma_{_{T}} = {\langle (\bec{\nabla} {\bf \cdot} {\bf u})^{2} \rangle \over \langle (\bec{\nabla} {\bf \times} {\bf u})^{2} \rangle} \equiv \sigma_u
\;,
 \label{AD9}
 \end{eqnarray}
where ${\bf u}$ is the fluid velocity field.
For a small yet finite correlation time (i.e., for a small Strouhal numbers, ${\rm Sr} = \tau_c \sqrt{\langle {\bf u}^{2} \rangle}/ \ell \ll 1$), the parameter $\sigma_{_{T}}$
can be estimated as
\begin{eqnarray}
\sigma_{_{T}} = \sigma_u + {2 \, {\rm Sr}^2\over 3} \, \Big(1+ {913 \, \sigma_u^2 \over 12 \, (1 +\sigma_u)} \Big) + O({\rm Sr}^4) \; .
 \label{AD10}
 \end{eqnarray}
For derivation of Eq.~(\ref{AD10}) we used Eq.~(C12) given in \cite{EKR02}. For inertial particles with small Stokes time the corrections $\sim O({\rm St}^2)$  in Eqs.~(\ref{AD7})-(\ref{AD9}) can be neglected (see subsection IV-B), where ${\rm St} = \tau_s / \tau_\eta$ is the Stokes number, $\tau_s$ is the Stokes time of inertial particles and $\tau_\eta$ is the Kolmogorov time scale.

Equation~(\ref{B2}) with $I({\bf R})=0$ has been derived in \cite{KR68} for a delta-correlated in time random incompressible $(b=0)$ velocity field. For a turbulent compressible $(b \not= 0)$ velocity field with a finite correlation time Eq.~(\ref{B2}) has been derived in \cite{EKR02} by means of stochastic calculus, i.e., Wiener path integral representation of the solution of the Cauchy problem for Eq.~(\ref{B1}), using Feynman-Kac formula and Cameron-Martin-Girsanov theorem. The comprehensive description of this approach can be found in \cite{EKR95,EKR02,ZRS90}.

\subsection{Gradient of the mean particle number density}

A nonzero gradient of the mean particle number density in stratified turbulence with external mean temperature gradient is caused by the phenomenon of turbulent thermal diffusion \cite{EKR96,EKR00,EEKR04,BEE04,EEKR06a,EEKR06b,SEKR09}.
This phenomenon in turbulent stratified flows results in the non-diffusive flux of particles in the direction of the heat flux. Particles are accumulated in the vicinity of the minimum of the mean temperature of the surrounding fluid. This causes formation of large-scale inhomogeneities in spatial distribution of particles. Turbulent thermal diffusion has been detected in two experimental set-ups: oscillating-grids turbulence generator \cite{EEKR04,BEE04,EEKR06a} and multi-fan turbulence generator  \cite{EEKR06b}. The experiments have been performed for stably and unstably stratified fluid. In these experiments, even with strongly inhomogeneous temperature fields, particles in turbulent fluid accumulate in the regions of temperature minima, in a good agreement with the theory of turbulent thermal diffusion \cite{EKR96,EKR00}.

Let us discuss the physics of the phenomenon of turbulent thermal diffusion. The velocity of particles, ${\bf v}$,  depends on the velocity of the surrounding fluid, ${\bf u}$, and it can be determined from the equation of motion for a particle.  When $\rho_p \gg \rho$, this equation represents a balance of particle inertia with the fluid drag force produced by the motion of the particle relative to the surrounding fluid, $d{\bf v}/dt = ({\bf u} - {\bf v}) / \tau_s$, where $\tau_s$ is the particle Stokes time, $\rho$ is the fluid density and $\rho_p$ is the material density of a particle. Solution of the equation of motion for small Stokes numbers, ${\rm St} \ll 1$, reads:
\begin{eqnarray}
{\bf v} = {\bf u} - {\rm St} \, \biggl[{\partial {\bf u}
\over \partial t} + ({\bf u} {\bf \cdot} \bec{\nabla}) {\bf u}
\biggr] + {\rm O}({\rm St}^2) \;,
\label{LBB2}
\end{eqnarray}
(see, e.g., \cite{M87}). This solution is written in dimensionless form, where the time is measured in the units of Kolmogorov time scales. The second term in Eq.~(\ref{LBB2}) describes the difference between the local fluid velocity and particle velocity arising due to the small but finite inertia of the particle. In this study we consider low Mach numbers turbulent flow with $\bec\nabla {\bf \cdot} \, {\bf u} = - \rho^{-1} \, ({\bf u} {\bf \cdot} \bec{\nabla}) \rho \not= 0$. Equation~(\ref{LBB2}) for the velocity of particles and Navier-Stokes equation for the fluid for large Reynolds numbers yield the equation for $\bec\nabla {\bf \cdot} \, {\bf v}$ written in dimensional form:
\begin{eqnarray}
\bec\nabla {\bf \cdot} \, {\bf v} &=& \bec\nabla {\bf \cdot} \, {\bf u}
- \tau_s \, \bec\nabla {\bf \cdot} \,  \biggl( {d{\bf u} \over dt} \biggr)
+ {\rm O}(\tau_s^2)
\nonumber \\
&=& - {1 \over \rho} \, ({\bf u} {\bf \cdot} \bec{\nabla}) \rho + {\tau_s \over \rho}  \,\bec{\nabla}^2 p  + {\rm O}(\tau_s^2) \;,
\label{LBB3}
\end{eqnarray}
where $p$ is the fluid pressure.

The physical mechanism of the phenomenon of turbulent thermal diffusion for inertial particles can be explained as follows. Due to inertia, particles inside the turbulent eddies drift out to the boundary regions between the eddies (the regions with the decreased velocity of the turbulent fluid flow). Neglecting non-stationarity and molecular viscosity, the estimate based on the Bernoulli's law implies that these are the regions with the increased pressure of the surrounding fluid. Consequently, particles are accumulated in the regions with the maximum pressure of the turbulent fluid. Indeed, due to the inertia effect $\bec\nabla {\bf \cdot} \, {\bf v} \propto (\tau_s /\rho) \,\bec{\nabla}^2 p \not=0$  even for incompressible fluid flow [see Eq.~(\ref{LBB3})]. On the other hand, for large Peclet numbers, when we can neglect the molecular diffusion of particles in Eq.~(\ref{B1}),  $\bec\nabla {\bf \cdot} \, {\bf v} \propto - d n / d t$. This implies that in regions with maximum pressure of turbulent fluid (i.e., where $\bec{\nabla}^2 p < 0)$ there is accumulation of inertial particles (i.e.,  $dn / dt \propto - (\tau_s /\rho) \,\bec{\nabla}^2 p > 0)$. Similarly, there is an outflow of inertial particles from regions with the minimum pressure of fluid.

In case of homogeneous and isotropic turbulence without external large-scale gradients of temperature, a drift from regions with increased (decreased) concentration of particles by a turbulent flow of fluid is equiprobable in all directions. Therefore pressure (temperature) of the fluid is not correlated with the turbulent velocity field and there exists only turbulent diffusion of particles.

Situation drastically changes in a turbulent fluid with a mean temperature gradient. In this case, the heat flux $\langle {\bf u} \, \theta \rangle$ is not zero, i.e., fluctuations of fluid temperature, $\theta$, and velocity are correlated. We consider low-Mach-number flows $({\cal M}= u/c_s \ll 1, \, c_s$ is the sound speed) and study mean-field effects. For low-Mach-number isothermal flows, the mean fluid mass flux $\langle {\bf u} \, \rho' \rangle$ is very small $(\sim O({\cal M}^2))$ (see, e.g., \cite{CH02}), i.e., the fluctuations of the fluid density $\rho'$ and velocity ${\bf u}$ are weakly correlated.
Moreover, in stratified turbulent flows with imposed mean temperature gradient the total mass current in the cell reference frame vanishes.

On the other hand, fluctuations of pressure must be correlated with the fluctuations of velocity due to a non-zero turbulent heat flux, $\langle {\bf u} \, \theta \rangle \not = 0$. Indeed, using the equation of state for an ideal gas we find that $p / P = \rho' / \rho + \theta / T$,
and  $\langle {\bf u} \, p \rangle /P = \langle {\bf u} \, \theta \rangle / T$, where $P$, $\,T$ and $\rho$ are the mean fluid pressure, temperature and density, respectively. Therefore, the fluctuations of temperature and pressure are correlated and the pressure fluctuations cause fluctuations of the number density of particles. The correlation between $p$ and $\theta$, necessary for $P \langle \theta \, {\bf u} \rangle =T \langle p \, {\bf u} \rangle$, arises from the buoyancy component of $p$ and from the effect of non-uniform mass density in the Navier-Stokes equation.

Increase (decrease) of the pressure of surrounding fluid is accompanied by accumulation (outflow) of the particles, respectively. The direction of the mean flux of particles coincides with the direction of the heat flux of temperature - towards the minimum of the mean temperature. Therefore, the particles are accumulated in this region (for more details, see \cite{EKR96}).

Equation for the evolution of the mean number density $N$ of particles reads:
\begin{eqnarray}
{\partial N \over \partial t} + \bec\nabla {\bf \cdot} \, \big[N \, {\bf V} + {\bf F}^{(n)} \big] =0 \;,
\label{LBB4}
\end{eqnarray}
where ${\bf V}$ is the mean particle velocity, ${\bf F}^{(n)} = \langle {\bf v} \, \Theta \rangle$ is the turbulent flux of particles that includes contributions of turbulent thermal diffusion and turbulent diffusion (see \cite{EKR96,EKR00}):
\begin{eqnarray}
{\bf F}^{(n)} = {\bf V}^{\rm eff} \, N - D_{_{T}} \, \bec\nabla  N \;,
\label{LBB5}
\end{eqnarray}
$D_{_{T}} \approx \ell_0 \,u_0$ is the coefficient of turbulent diffusion, $u_0$ is the characteristic turbulent velocity in the maximum scale $\ell_0$ of turbulent motions,  ${\bf V}^{\rm eff}$ is the effective velocity caused by turbulent thermal diffusion is given by the following equation:
\begin{eqnarray}
{\bf V}^{\rm eff} = - \tau \, \langle {\bf v} \, (\bec\nabla {\bf \cdot} \, {\bf v}) \rangle \; .
\label{NB6}
\end{eqnarray}
Equation~(\ref{NB6}) for the effective velocity has been derived using different rigorous methods in \cite{EKR96,EKR00,PM02,RE05,SEKR09}. Note that even a simple dimensional analysis yields the estimate for the effective velocity ${\bf V}^{\rm eff}$ that coincides with Eq.~(\ref{NB6}). Indeed, the magnitude of $\partial \Theta / \partial t + \bec\nabla {\bf \cdot} \, {\bf Q} - D_m \bec{\nabla}^2 \Theta$ in Eq.~(\ref{RB1}) can be estimated as $\Theta/\tau$. Therefore, the turbulent component $\Theta$  of particle number density is of the order of  $\Theta \approx - \tau \, \bec\nabla {\bf \cdot} \, (N \, {\bf v}) = - \tau \, [N (\bec\nabla {\bf \cdot} \, {\bf v}) +  ({\bf v} {\bf \cdot} \bec{\nabla}) N]$. Now let us determine the turbulent flux of particles $F_i^{(n)} = \langle v_i \, \Theta \rangle$:
\begin{eqnarray}
F_i^{(n)} = - N \, \tau \, \langle v_i \, (\bec\nabla {\bf \cdot} \, {\bf v}) \rangle  - \tau \, \langle v_i v_j \rangle \nabla_j N \;,
\label{NBB8}
\end{eqnarray}
where the first term in the right hand side of Eq.~(\ref{NBB8}) determines the turbulent flux of particles caused by turbulent thermal diffusion:  $- N \, \tau \, \langle v_i \, (\bec\nabla {\bf \cdot} \, {\bf v}) \rangle = V^{\rm eff}_i \, N$, while the second term in the right hand side of Eq.~(\ref{NBB8}) determines the turbulent flux of particles caused by turbulent diffusion: $\tau \, \langle v_i v_j \rangle \nabla_j N = D_{_{T}} \,\nabla_i N$. In the latter estimate we neglected the anisotropy of turbulence for simplicity.
More detailed analysis shows that the effective velocity caused by turbulent thermal diffusion is given by:
\begin{eqnarray}
{\bf V}^{\rm eff} = - \alpha \, D_{_{T}} \, {\bec\nabla T \over T}\;,
\label{LBB9}
\end{eqnarray}
where ${\bf V}^{\rm eff} = {\bf U}({\bf R}=0)$ (see subsection IV-C). The turbulent thermal diffusion ratio, $\alpha$, is given by the following equation:
\begin{eqnarray}
\alpha = 1 + {\gamma \, W_g \, L_P \, \ln({\rm Re}) \over u_0 \, \ell_0} \;,
\label{LBB16}
\end{eqnarray}
(see \cite{EKR96,EKR00}), where $\gamma=c_{\rm p}/c_{\rm v}$ is the ratio of specific heats, ${\bf W}_{g} = \tau_s \, {\bf g}$  is the terminal fall velocity of particles, $\tau_s = m_p / 6 \pi \rho \, \nu a_p$ is the Stokes time for small spherical particles of the radius $a_p$ and mass $m_p$, $\, {\bf g}$ is the acceleration of gravity, $L_P^{-1} = |\nabla_z P / P|$ and ${\rm Re}= u_0 \ell_0 / \nu$ is the Reynolds number. For gases and non-inertial particles $\alpha =1$. For derivation of Eqs.~(\ref{LBB9}) and~(\ref{LBB16}) we took into account the equation of state, neglected the mass flux of fluid, $\langle {\bf u} \, \rho' \rangle$, for the low-Mach-number flows, and used the identity $\tau_s = \rho \, W_g \, L_P / P$, where $|\nabla_z P| = \rho \, g$. The steady-state solution of Eq.~(\ref{LBB4}) at ${\bf V}=0$ reads:
\begin{eqnarray}
{\bec\nabla N \over N} = - \alpha \, {\bec\nabla T \over T} \; .
\label{BB15}
\end{eqnarray}
We will use Eq.~(\ref{BB15}) in order to determine the source function $I({\bf R})$.

\subsection{Functions $B({\bf R})$ and ${\bf U}^{(S)}({\bf R})$}

Let us consider the case $\alpha^2 \gg 1$. The magnitude of $\alpha^2$ in the experiment was of the order of $10$ - $15$. Note that it is not easy to determine the exact value of $\alpha^2$ in the experiments because of a size distribution of particles. In particular, the particle size in the experiments varies from 3 to 40 $\mu$m, with the mean value 10 $\mu$m. On the other hand, in the analysis of the particle clustering we used a threshold in the light intensity. This implies that we did not take into account very small particles in the data analysis of the particle clustering.

In this case $\bec\nabla {\bf \cdot} \, {\bf v} \approx (\tau_s/\rho) \, \bec{\nabla}^2 p$, and the functions $B({\bf R})$ determined by Eq.~(\ref{AL3}), is given by the following equation:
\begin{widetext}
\begin{eqnarray}
B({\bf R}) &\approx& {2 \tau_s^2 \over \rho^2} \, \langle \tau \big[\bec{\nabla}^2 p({\bf x}) \big]\, \bec{\nabla}^2 p({\bf y}) \rangle \approx
{2 \tau_s^2 \over \rho^2} \, \Big[ {P^2 \over T^2} \, \langle \tau \big[\bec{\nabla}^2 \theta({\bf x})\big] \, \bec{\nabla}^2 \theta({\bf y}) \rangle
+ {P^2 \over \rho^2} \, \langle \tau \big[\bec{\nabla}^2 \rho'({\bf x})\big] \, \bec{\nabla}^2 \rho'({\bf y}) \rangle
\nonumber\\
&& + {P^2 \over \rho \, T} \, \Big(\langle \tau \big[\bec{\nabla}^2 \rho'({\bf x})\big] \, \bec{\nabla}^2 \theta({\bf y}) \rangle
+ \langle \tau \big[\bec{\nabla}^2 \theta({\bf x})\big] \, \bec{\nabla}^2 \rho'({\bf y}) \rangle \Big) \Big] \;,
\label{AL5}
\end{eqnarray}
\end{widetext}
\noindent
where $\bec{\nabla}^2 p({\bf x}) = \big[\bec\nabla^{({\bf x})} \big]^2 p({\bf x})$. Hereafter we omit the argument $t$ in the correlation function.
In derivation of this equation we used the relationship
\begin{eqnarray}
{p \over P} ={\rho' \over \rho} + {\theta \over T} + O(\rho' \, \theta)
\;,
\label{AD1}
\end{eqnarray}
that follows from the equation of state for ideal gas. We also take into account that characteristic spatial scales for fluctuations of fluid pressure, temperature and density are much less than those for the mean fields.

In turbulence with imposed turbulent heat flux (e.g., the imposed mean temperature gradient in the oscillating grid turbulence), the correlation function $\langle \big[\bec{\nabla}^2 \theta({\bf x})\big] \, \bec{\nabla}^2 \theta({\bf y}) \rangle$ is much larger than the correlation functions of density-density fluctuations or density-temperature fluctuations, i.e.,
\begin{eqnarray}
{1 \over T^2} \, |\langle \big[\bec{\nabla}^2 \theta({\bf x})\big] \, \bec{\nabla}^2 \theta({\bf y}) \rangle | \gg {1 \over \rho^2} \, |\langle \big[\bec{\nabla}^2 \rho'({\bf x})\big] \, \bec{\nabla}^2 \rho'({\bf y}) \rangle | \;,
\nonumber\\
\label{AD2}\\
{1 \over T} \, |\langle \big[\bec{\nabla}^2 \theta({\bf x})\big] \, \bec{\nabla}^2 \theta({\bf y}) \rangle | \gg {1 \over \rho} \, |\langle \big[\bec{\nabla}^2 \rho'({\bf x})\big] \, \bec{\nabla}^2 \theta({\bf y}) \rangle | \;,
\nonumber\\
\label{AD3}\\
{1 \over T} \, |\langle \big[\bec{\nabla}^2 \theta({\bf x})\big] \, \bec{\nabla}^2 \theta({\bf y}) \rangle | \gg {1 \over \rho} \, |\langle \big[\bec{\nabla}^2 \theta({\bf x})\big] \, \bec{\nabla}^2 \rho'({\bf y}) \rangle | \; .
\nonumber\\
\label{AD4}
\end{eqnarray}
Indeed, the correlation function $\langle \big[\bec{\nabla}^2 \theta({\bf x})\big] \, \bec{\nabla}^2 \theta({\bf y}) \rangle$ is caused by the turbulent heat flux, i.e., $\langle \theta({\bf x}) \, \theta({\bf y}) \rangle \propto - \tau_0  \, \langle u_i({\bf x}) \, \theta({\bf y}) \rangle \, (\nabla_i T)$ [see Eqs.~(\ref{BD2}) and~(\ref{BD4})], where $\tau_0$ is the characteristic turbulent time. On the other hand, the correlation functions of density-density fluctuations or density-temperature fluctuations are nearly independent of the turbulent heat flux, and they are proportional to the mass flux $\langle {\bf u}({\bf x}) \, \rho'({\bf y}) \rangle$, which is very small.

In particular, the temperature fluctuations can be estimated as $\theta \propto - \tau_0 \, u_i \, \nabla_i T$. Then the temperature-density fluctuations are estimated as $\langle \theta({\bf x}) \, \rho'({\bf y}) \rangle \propto - \tau_0  \, \langle u_i({\bf x}) \, \rho'({\bf y}) \rangle \, (\nabla_i T)$. The density fluctuations are determined by the continuity equation:
\begin{eqnarray}
{\partial \rho' \over \partial t} = - \bec{\nabla} {\bf \cdot}  (\rho {\bf u}) + O(\rho' {\bf u}) \;,
\label{AD5}
\end{eqnarray}
and the correlation function of density-density fluctuations $\langle \rho'({\bf x}) \, \rho'({\bf y}) \rangle$ is determined by the following equation
\begin{eqnarray}
&& {\partial \over \partial t} \langle \rho'({\bf x}) \, \rho'({\bf y}) \rangle =  - \rho \, \big[\nabla_i^{(y)} \, \langle \rho'({\bf x}) \, u_i({\bf y}) \rangle
\nonumber\\
&&\quad + \nabla_i^{(x)} \, \langle \rho'({\bf y}) \, u_i({\bf x}) \rangle \big] -{\nabla_i \, \rho \over \rho} \, \big[\langle \rho'({\bf x}) \, u_i({\bf y}) \rangle
\nonumber\\
&&\quad + \langle \rho'({\bf y}) \, u_i({\bf x}) \rangle \big] \;,
\label{AD6}
\end{eqnarray}
which follows from Eq.~(\ref{AD5}).
Since $\langle \rho'({\bf x}) \, u_i({\bf y}) \rangle$ is very small, and is nearly independent of the turbulent heat flux, the correlation functions of the density-density fluctuations or density-temperature fluctuations are much smaller than the correlation functions of the temperature-temperature fluctuations.

In ${\bf k}$ space the correlation function $\langle \tau \big[\bec{\nabla}^2 \theta({\bf x})\big] \, \big[\bec{\nabla}^2 \theta({\bf y})\big] \rangle$ reads:
\begin{eqnarray}
\langle \tau \big[\bec{\nabla}^2 \theta({\bf x})\big] \, \big[\bec{\nabla}^2 \theta({\bf y})\big] \rangle &=& \int \tau(k) \, k^4 \,  \langle \theta({\bf k}) \, \theta(-{\bf k}) \rangle
\nonumber\\
&& \times \, \exp \big(i {\bf k} {\bf \cdot} {\bf R}\big) \, d{\bf k} .
\label{AL6}
\end{eqnarray}
Similarly, the function ${\bf U}^{(S)}({\bf R})$ determined by Eq.~(\ref{AL4}), is given by the following equation:
\begin{eqnarray}
{\bf U}^{(S)}({\bf R}) &\approx& - {2 \tau_s \over \rho} \, \langle \tau {\bf u}({\bf x}) \, \big[\bec{\nabla}^2 p({\bf y}) \big]\rangle
\nonumber\\
&\approx& - {2 \tau_s \, P \over \rho \, T} \, \langle \tau {\bf u}({\bf x})\, \big[\bec{\nabla}^2 \theta({\bf y})\big] \rangle \; .
\label{AL7}
\end{eqnarray}
In ${\bf k}$ space the correlation function $\langle \tau {\bf u}({\bf x})\, \big[\bec{\nabla}^2 \theta({\bf y})\big] \rangle$ reads:
\begin{eqnarray}
\langle \tau {\bf u}({\bf x})\, \big[\bec{\nabla}^2 \theta({\bf y})\big] \rangle &=& - \int \tau(k) \, k^2 \,  \langle {\bf u}({\bf k}) \, \theta(-{\bf k}) \rangle
\nonumber\\
&& \times \, \exp \big(i {\bf k} {\bf \cdot} {\bf R}\big) \, d{\bf k}  \; .
\label{AL8}
\end{eqnarray}

In order to determine the correlation functions $\langle \theta({\bf k}) \, \theta(-{\bf k}) \rangle$ and $\langle {\bf u}({\bf k}) \, \theta(-{\bf k}) \rangle$ we use the following evolutionary equation for the temperature field $T_{\rm tot}(t, {\bf r})$ in a turbulent flow:
\begin{eqnarray}
{\partial T_{\rm tot} \over \partial t} + ({\bf u} {\bf \cdot} \bec\nabla) \, T_{\rm tot} + (\gamma - 1) \, ({\bf  \nabla} {\bf \cdot} {\bf u}) \, T_{\rm tot} = D \,\bec{\nabla}^2 T_{\rm tot} \;,
\nonumber\\
\label{ABB1}
\end{eqnarray}
where $D$ is the coefficient of molecular temperature diffusion, $\gamma=c_{\rm p}/c_{\rm v}$ is the ratio of specific heats, ${\bf u}$ is the fluid velocity field that satisfies to continuity equation in anelastic approximation for a low-Mach-number flow:
\begin{eqnarray}
{\bf  \nabla} {\bf \cdot} (\rho \, {\bf u}) =0 \; .
\label{AB1}
\end{eqnarray}
Combining Eqs.~(\ref{ABB1}) and~(\ref{AB1}) we obtain the following equation:
\begin{eqnarray}
{\partial T_{\rm tot} \over \partial t} + (\hat{\bf u} {\bf \cdot} \bec\nabla) \, T_{\rm tot} = D \,\bec{\nabla}^2 T_{\rm tot} \;,
\nonumber\\
\label{ALBB1}
\end{eqnarray}
where $\hat{\bf u}= \gamma {\bf u}$.
Averaging Eq.~(\ref{ALBB1}) over an ensemble of turbulent velocity field we obtain the equation for the evolution of the mean temperature field $T(t, {\bf r})$:
\begin{eqnarray}
{\partial T \over \partial t} + \bec\nabla {\bf \cdot} \, {\bf F} = D \,\bec{\nabla}^2 T \;,
\label{ABB4}
\end{eqnarray}
where ${\bf F} = \langle \hat{\bf u} \, \theta \rangle$ is the heat flux, and for simplicity we consider the fluid flow with a zero mean velocity. Note that for a low-Mach-number flow without imposed external pressure gradient $\bec\nabla \rho/\rho \approx - \bec\nabla T /T$.

Subtracting Eq.~(\ref{ABB4}) from Eq.~(\ref{ALBB1}) yields equation for the temperature fluctuations:
\begin{eqnarray}
{\partial \theta \over \partial t} + Q - D \bec{\nabla}^2 \theta  = I \;,
\label{AC1}
\end{eqnarray}
where $I = - (\hat{\bf u} {\bf \cdot} \bec{\nabla}) T$ is the source term and $Q = \bec\nabla {\bf \cdot} \, [\hat{\bf u} \, \theta - \langle \hat{\bf u} \,\theta \rangle] + \theta (\hat{\bf u} {\bf \cdot} \bec\nabla) \rho/ \rho$  is the nonlinear term. In Eq.~(\ref{ABB4}) we neglected the term $\langle \theta \, \hat{\bf u} \rangle {\bf \cdot} \bec\nabla \rho/ \rho$ which is quadratic in large-scale spatial derivatives.

Now we derive formulae for the function $F_i(t, {\bf k}) = \langle \hat u_i(t, {\bf k}) \, \theta(t, -{\bf k}) \rangle$ and the temperature fluctuation function $E_\theta(t, {\bf k}) = \langle \theta(t, {\bf k}) \, \theta(t, -{\bf k}) \rangle$ using the $\tau$ approach that is valid for large Peclet and Reynolds numbers. Using Eq.~(\ref{AC1}) written in a Fourier space we derive equation for the instantaneous two-point second-order correlation functions:
\begin{eqnarray}
{dF_i \over dt} = \langle \hat u_i({\bf k}) \, I(-{\bf k}) \rangle + \hat{\cal M} F_i^{(III)}({\bf k}) \;,
\label{BD1}\\
{dE_\theta \over dt} = 2 \langle \theta({\bf k}) \, I(-{\bf k}) \rangle + \hat{\cal M} E_\theta^{(III)}({\bf k}) \;,
\label{BD2}
\end{eqnarray}
where $\hat{\cal M} F_i^{(III)}({\bf k}) = - [\langle \hat u_i \, Q \rangle + \langle (\partial \hat u_i / \partial t) \, \theta \rangle - D \langle \hat u_i \, \bec{\nabla}^2 \theta \rangle]_{\bf k}$ and $\hat{\cal M} E_\theta^{(III)}({\bf k}) = - [\langle \theta \, Q \rangle - D \langle \theta \, \bec{\nabla}^2 \theta \rangle]_{\bf k}$ are the third-order moment terms appearing due to the nonlinear terms which include also molecular diffusion term.

The equation for the second moment includes the first-order spatial
differential operators $\hat{\cal M}$  applied to the third-order
moments $F^{(III)}$. A problem arises how to close the system, i.e.,
how to express the third-order terms $\hat{\cal M}
F^{(III)}$ through the lower moments $F^{(II)}$
(see, e.g., \cite{O70,MY75,Mc90}). We use the spectral $\tau$ approximation which postulates that the deviations of the third-moment terms, $\hat{\cal M} F^{(III)}({\bf k})$, from the contributions to these terms afforded by the background turbulence, $\hat{\cal M} F^{(III,0)}({\bf k})$, can be expressed through the similar deviations of the second moments, $F^{(II)}({\bf k}) - F^{(II,0)}({\bf k})$:
\begin{eqnarray}
&& \hat{\cal M} F^{(III)}({\bf k}) - \hat{\cal M} F^{(III,0)}({\bf
k})
\nonumber\\
&& \quad \quad\quad = - {1 \over \tau_r(k)} \, \Big[F^{(II)}({\bf k}) - F^{(II,0)}({\bf k})\Big] ,
\label{AD2}
\end{eqnarray}
(see, e.g., \cite{O70,PFL76,RK07}), where $\tau_r(k)$ is the scale-dependent relaxation time, which can be identified with the correlation time $\tau(k)$ of the turbulent velocity field for large Reynolds and Peclet numbers. The functions with the superscript $(0)$ correspond to the background turbulence with a zero gradient of the mean temperature. Validation of the $\tau$ approximation for different situations has been performed in numerous numerical simulations and analytical studies (see, e.g., review \cite{BS05}; and also discussion in \cite{RK07}, Sec. 6).

Note that the contributions of the terms with the superscript $(0)$ vanish because when the gradient of the mean temperature is zero, the turbulent heat flux and the temperature fluctuations vanish. Consequently, Eq.~(\ref{AD2}) for $\hat{\cal M} F_i^{(III)}({\bf k})$ reduces to $\hat{\cal M} F_i^{(III)}({\bf k}) = - F_i({\bf k}) / \tau(k)$ and $\hat{\cal M} E_\theta^{(III)}({\bf k}) = - E_\theta({\bf k}) / \tau(k)$.
We also assume that the characteristic time of variation of the second moments $F_i({\bf k})$ and $E_\theta({\bf k})$ are substantially larger than the correlation time $\tau(k)$ for all turbulence scales. Therefore, in a steady-state Eqs.~(\ref{BD1}) and~(\ref{BD2}) yield the following formulae for the functions $E_\theta({\bf k})$ and $F_i({\bf k})$:
\begin{eqnarray}
\langle \theta({\bf k}) \, \theta(-{\bf k}) \rangle &=& 2 \tau^2(k) \langle \hat u_i({\bf k}) \, \hat u_j(-{\bf k}) \rangle \, (\nabla_i T) \, (\nabla_j T) \;,
\nonumber\\
\label{BD3}\\
\langle \hat u_i({\bf k}) \, \theta(-{\bf k}) \rangle &=& - \tau(k) \langle \hat u_i({\bf k}) \, \hat u_j(-{\bf k}) \rangle \, \nabla_j T  \; .
\label{BD4}
\end{eqnarray}
Now we use the Kolmogorov model for the turbulent correlation time, $\tau(k) = 2 \, \tau_0 \, (k / k_{0})^{-2/3}$, where $\tau_0 = \ell_0 / u_{0}$ is the characteristic turbulent time, the wave number $k_{0} = 1 / \ell_0$, the length $\ell_0$ is the maximum scale of random motions and $u_0$ is the characteristic velocity in the maximum scale of random motions. We also take into account that $\langle u_i({\bf k}) \, \theta(-{\bf k}) \rangle = \gamma^{-1} \langle \hat u_i({\bf k}) \, \theta(-{\bf k}) \rangle$. Substituting Eqs.~(\ref{BD3}) and~(\ref{BD4}) into Eqs.~(\ref{AL6}) and~(\ref{AL8}), respectively, and using Eqs.~(\ref{AL5}) and~(\ref{AL7}) we arrive at the following expressions for the functions $B({\bf R})$ and ${\bf U}^{(S)}({\bf R})$:
\begin{eqnarray}
B({\bf R}) &=& - 6 I_\ast \, \tau_0 \, \Delta \, u_{zz}({\bf R}) ,
\label{AL10}\\
U_i^{(S)}({\bf R}) &=& - \alpha \, D_{ij}^{^{T}}({\bf R}) \, {\nabla_j T \over T} ,
\label{AL11}
\end{eqnarray}
where $u_{zz}({\bf R}) = \langle u_{z}({\bf x}) \, u_{z}({\bf x}+{\bf R}) \rangle$. In derivation of Eqs.~(\ref{AL10}) and~(\ref{AL11}) we take into account the following model for the second moments of turbulent velocity field, $\langle u_i({\bf k}) \, u_j(-{\bf k}) \rangle$ in ${\bf k}$ space:
\begin{eqnarray}
&& \langle u_i({\bf k}) \, u_j(-{\bf k}) \rangle = {u_0^2 \, E(k) \over 8 \pi k^2} \Big[\delta_{ij} - {k_i \, k_j \over k^2} \Big] \;,
\label{ALD3}
\end{eqnarray}
where the energy spectrum function is $E(k) = (2/3)\, k_0^{-1} \, (k / k_{0})^{-5/3}$. In derivation of Eq.~(\ref{AL10}) we also used the following identity:
\begin{eqnarray}
\int \tau^3(k) \, k^4 \,  E({\bf k}) \, \exp \big(i {\bf k} {\bf \cdot} {\bf R}\big) \, d{\bf k} = - 8 \tau_0 \, \Delta \, E({\bf R}) .
\label{AL12}
\end{eqnarray}
The theoretical study performed in this subsection allows us to determine the source function $I({\bf R})$ determined by Eq.~(\ref{IB1}).

\subsection{Equation for two-point correlation function of particle number density}

The theoretical analysis performed in previous subsection has shown that the source function $I({\bf R})$ determined by Eq.~(\ref{IB1}), can be rewritten in the following form:
\begin{eqnarray}
I({\bf R}) &=& I_\ast \Big[{D_{zz}^{^{T}}({\bf R}) \over D_{_{T}}}  \,\ln^2 \, {\rm Re}  - 6 \tau_0^2 \Delta \, u_{zz}({\bf R}) \Big] \, {N^2 \over \tau_0} ,
\nonumber\\
\label{IB2}\\
I_\ast &=& {4 \alpha^2 \over 3 \ln^2 \, {\rm Re}} \, \Big({ \ell_0 \, \nabla_z T \over T} \Big)^2 ,
\label{IB3}
\end{eqnarray}
where $u_{zz}({\bf R}) = \langle u_{z}({\bf x}) \, u_{z}({\bf x}+{\bf R}) \rangle$ and we have considered the case $\alpha^2 \gg 1$.  In derivation of Eq.~(\ref{IB2}) we used Eqs.~(\ref{AL10}) and~(\ref{AL11}).

Let us discuss the assumptions underlying the employed model of particle transport in turbulent flow. We use the tensor of turbulent diffusion $D^{^{\rm T}}_{ij} ({\bf R})$ for isotropic and homogeneous turbulent flow. In our experiments the velocity field is weakly anisotropic.
A main contribution to the magnitude of fluctuations of particle number density is due to the mode with the minimum damping rate \cite{EKR95,EEKR09}. This mode is an isotropic solution of Eq.~(\ref{B2}). Consequently, it is plausible to neglect the anisotropic effects. This assumption is also supported by our measurements of the two-point second-order correlation function $\Phi(t,{\bf R})$. The turbulence parameters $\ell_0$ and $u_0$ vary slowly in the probed region.

Note that the mechanism of mixing related to the tangling of the gradient of the mean particle number density is quite robust. The properties of the tangling are not very sensitive to the exponent of the energy spectrum of the background turbulence. The requirements that turbulence should be isotropic, homogeneous, and should have a very long inertial range (a fully developed turbulence), are not necessary for the tangling mechanism. Anisotropy effects can complicate the theoretical analysis, but do not introduce new physics in the clustering process. The reason is that the main contribution to the tangling clustering is at Kolmogorov (viscous) scale of turbulent motions. At this scale turbulence can be considered as nearly isotropic, while anisotropy effects can be essential in the vicinity of the maximum scale of turbulent motions.

Using these arguments, we consider the tensor $D^{^{\rm T}}_{ij} ({\bf R})$ in the following form:
\begin{eqnarray}
D^{^{\rm T}}_{ij} ({\bf R}) &=& D_{_{\rm T}} \, \Big[ [F(R) + F_c(R)] \delta_{ij} + R F'_c \, {R_i R_j \over R^2}
\nonumber\\
&& + {R F' \over 2} \Big(\delta_{ij} - {R_i R_j \over R^2} \Big) \Big]
\;,
\label{T15}
\end{eqnarray}
where $D_{_{T}} = \ell_0 \,u_0/3$, $\, F(0) = 1 - F_c(0)$ and $F'=dF/dR$. The function $F_c(R)$ describes the compressible (potential) component, whereas $F(R)$ corresponds to vortical (incompressible) part of the turbulent diffusion tensor.

Now let us study a zero-mode (i.e., a mode with $\partial \Phi / \partial t = 0)$. Using Eqs.~(\ref{B2}) and~(\ref{T15}) we derive equation for the two-point second-order correlation function $\Phi(R)$ written in a dimensionless form:
\begin{eqnarray}
{1 \over M(R)} \Big[\Phi'' + 2 \, \Big({1 \over R} + \chi(R) \Big) \, \Phi' \Big] + B(R) \, \Phi = - I(R) ,
\label{T1}
\end{eqnarray}
where time $t$ is measured in units of $\tau_0=\ell_0/u_0$, distance
$R$ is measured in units of $\ell_0$, and
\begin{eqnarray}
{1 \over M(R)} &=& {2 \over {\rm Pe}} + {2 \over 3} [1 - F - (R F_c)']\;,
\label{L1}\\
\chi(R) &=&  - {M(R) \over 3} (F - 2 F_c)' \;,
\label{L2}\\
I(R) &=& I_\ast \Big[\ln^2 \, {\rm Re} \, \Big(1 - {3 \over 2 M(R)} - {R \, \chi(R) \over M(R)}\Big)
\nonumber\\
&& - 2 \Delta \, \Big(F_u(R) + {R F'_u(R) \over 3} \Big) \Big] ,
\label{IB4}
\end{eqnarray}
and ${\rm Pe} = u_0 \ell_0 / D_m$ is the Peclet number and $F_u(R)$ is the longitudinal correlation function of particle velocity field. In derivation of Eq.~(\ref{T1}) we have taken into account that for large $\alpha$ the term $|U^{(A)}({\bf R}) \Phi'| \ll |B(R) \, \Phi|$ for all scales.

The two-point correlation function $\Phi(R)$ satisfies the following boundary conditions: $\Phi'(R=0) = 0$ and $\Phi(R \to \infty) = 0$. This function has a global maximum at $R=0$ and therefore it satisfies the conditions:
\begin{eqnarray*}
\Phi''(R=0) < 0\,, \quad  \Phi(R=0) >  |\Phi (R>0)| \; .
\end{eqnarray*}

Particular formulas for the functions $M(R)$, $\, \chi(R)$ and $I(R)$ depend on the functions $F(R)$, $\, F_c(R)$ and $F_u(R)$. For instance, we may choose these functions in the following form:
\begin{eqnarray}
F(R) &=& {1 \over 1 +\sigma_{_{T}}} \, \exp[- f(R)] \;,
\label{C10}\\
F_c(R) &=& {\sigma_{_{T}} \over 1 +\sigma_{_{T}}} \, \exp[- a_c \, f(R)] \;,
\label{C11}\\
f(R) &=& {R^2 \over {\rm Re}^{-1/2} + R^{2/3}}  \;,
\label{C12}\\
F_u(R) &=& \exp\Big[- {R^2 \over {\rm Re}^{-1/2} + R^{4/3}}\Big] \; .
\label{IB5}
\end{eqnarray}
Equation~(\ref{C12}) is similar to the interpolation formula derived by Batchelor for the correlation function of the velocity field that is valid for a turbulence with Kolmogorov spectrum in the inertial range and for random motions in the viscous range of scales (see, e.g., \cite{MY75,Mc90}). In particular, in the inertial range of turbulent scales, ${\rm Re}^{-3/4} \ll R \ll 1$, the correlation function for the turbulent diffusion tensor is $F(R) \propto 1 - R^{4/3}$, where $R$ is measured in units of maximum scale of turbulent motions $\ell_0$. The corresponding correlation function for the turbulent velocity field $F_u(R) \propto 1 - R^{2/3}$. The difference between the scalings of the turbulent diffusion tensor and the correlation function of the turbulent velocity field is caused by the scaling of correlation time $\tau(R) \propto R^{2/3}$. On the other hand, in the viscous range, $R \ll {\rm Re}^{-3/4}$, the correlation function for the turbulent diffusion  tensor is $F(R) \propto 1 - {\rm Re}^{1/2} R^{2}$. This correlation function is similar to that for the velocity field because in the viscous range the correlation time is independent of scale. On the other hand, for large scales $R \gg 1$ there is no turbulence, so that for $R > 1$ the functions  $F(R)$ and $F_c(R)$ should sharply decrease to zero.

The particular choice of the functions $F(R)$, $\, F_c(R)$, $\, F_u(R)$ and $f(R)$ in the paper describes the well-known properties of turbulent velocity field obtained from laboratory experiments, numerical simulations and theoretical studies (see, e.g., \cite{MY75,Mc90}). In this study we choose the exponential form for the functions $F(R)$, $\, F_c(R)$ and $F_u(R)$. The final results are not sensitive to the form for these functions at large scales $R \gg 1$. Note also that the particular choice of the function $f(R)$ is also not very  important. It should describe turbulent motions in the inertial range (e.g., a turbulence with Kolmogorov spectrum in the inertial range) and random motions in the viscous range of scales. The parameter $a_c$ in Eq.~(\ref{C11}) characterizes the different spatial scalings of the compressible and incompressible parts of the tensor of the scale-dependent turbulent diffusion. Our analysis has shown that the two-point second-order correlation function of the particle number density is weakly dependent on the parameter $a_c$.

Analysis performed by Kraichnan in \cite{KR68} for the delta-correlated in time turbulent velocity field, showed that the characteristic damping rate of the particle number density fluctuations is very high $\gamma \sim \tau_\eta^{-1}$. The latter implies that the level of these fluctuations is very low $\sim (\gamma \,\tau_0)^{-1} \sim {\rm Re}^{- 1/2}$. In a real flow with a finite correlation time the degree of compressibility of the turbulent diffusion tensor $\sigma_{_{T}} \not = 0$, the characteristic damping rate of the fluctuations of the particle number density is not high $\gamma \leq \tau_0^{-1}$, and, therefore, the level of these fluctuations is not small.

\subsection{Asymptotic analysis and numerical solution}

Let us perform the detailed asymptotic analysis of the solution of Eq.~(\ref{T1}) for the two-point second-order correlation function $\Phi(R)$ in different ranges of scales. There are several characteristic regions in the solution for the correlation function $\Phi(R)$: (i) the viscous range $0 \leq R < {\rm Re}^{-3/4}$; (ii) the inertial range ${\rm Re}^{-3/4} \leq R \leq 1$ and (iii) the large scales $R \gg 1$ whereby there is no turbulence. Here $R$ is measured in units of the maximum scale of turbulent motions $\ell_0$. The dimensionless two-point second-order correlation function $\Phi(R)$ is determined by Eq.~(\ref{T1}). Equation~(\ref{T1}) in the viscous range $a_p \leq R < {\rm Re}^{-3/4}$ reads:
 \begin{eqnarray}
R^2 \Phi'' + 2R \Phi' + B_0 \Phi = - B_0 \;,
 \label{T2}
 \end{eqnarray}
where we take into account that in the viscous range of scales the functions $M(R)$, $\, \chi(R)$ and $I(R)$ are given by the following formulas:
 \begin{eqnarray}
{1 \over M(R)} &=& {2 \over {\rm Pe}} \, (1 + \beta_M \, {\rm Pe} \, \sqrt{\rm Re} \, R^2) \;,
 \label{IT1}\\
\chi(R) &=& - {2 \beta_\chi \over 3} \, M(R) \,  {\rm Re}^{1/4} R \;,
 \label{IT2}\\
I(R) &\approx& B_0 = 20 I_\ast {\rm Re} \;,
 \label{IT3}\\
\beta_\chi &=& {2 \sigma -1 \over 1 +\sigma_{_{T}}} \;, \quad \beta_M = {1 + 3 \sigma \over 3(1 +\sigma_{_{T}})} \;,
 \label{IT4}
 \end{eqnarray}
and $\sigma = a_c \, \sigma_{_{T}}$. Here we use asymptotics of the functions $F(R)$, $F_c(R)$ and $F_u(R)$ determined for $R \ll {\rm Re}^{-3/4}$ [see Eqs.~(\ref{C10})-(\ref{IB5})]. The solution of Eq.~(\ref{T2}) reads
\begin{eqnarray}
\Phi(R) = A_1 \Big({R \over a_p}\Big)^{-1/2} \cos \Big[\sqrt{B_0} \, \ln \, {R \over a_p} + \varphi_1 \Big] - 1 \; .
\label{T3}
\end{eqnarray}
The condition $\Phi'(R=0) = 0$ yields $\varphi_1 = - \arctan (1/ 2 \sqrt{B_0}) + \pi k$.

In the inertial range ${\rm Re}^{-3/4} \leq R \leq 1$, Eq.~(\ref{T1}) reads
\begin{eqnarray}
y^2 \tilde \Phi''(y) + 4y \, \tilde \Phi'(y) + \beta \tilde \Phi(y) = 0 \;,
\label{T10}
\end{eqnarray}
where $y = R^{1/3}$, the function $\tilde \Phi = \Phi + 1$, $\, \beta = 220 \,I_\ast / (9 \tilde \beta_M)$ and we take into account that in the inertial range the functions $M(R)$, $\, \chi(R)$ and $I(R)$ are given by the following formulas:
 \begin{eqnarray}
{1 \over M(R)} &=& \tilde \beta_M \, R^{4/3} \;,
 \label{IT5}\\
\chi(R) &=& - {4 \beta_\chi \over 9} \, M(R) \,  R^{1/3} \;,
 \label{IT6}\\
I(R) &\approx& B_0 = {220 \, I_\ast \over 81} \, R^{-4/3} \;,
 \label{IT7}\\
\tilde \beta_M &=& {2 (3 + 7 \sigma) \over 9(1 +\sigma_{_{T}})}  \; .
 \label{IT8}
 \end{eqnarray}
Equation~(\ref{T10}) has the following solution:
\begin{eqnarray}
\Phi(R) = A_2 \Big({R \over \ell_\eta}\Big)^{-1/3} \cos \Big[\sqrt{\beta} \, \Big({R \over \ell_\eta}\Big)^{-1/3} + \varphi_2 \Big] - 1 \;,
\nonumber\\
\label{T5}
\end{eqnarray}
where $\ell_\eta$ is the Kolmogorov length scale. For large scales, $R > 1$, there is no turbulence. The condition $\Phi(R) \to 0$ at $R\to 1$ yields: $A_2 \approx {\rm Re}^{1/4}$. Matching the functions $\Phi(R)$ and $\Phi'(R) $ at the boundary of the above-mentioned regions, i.e., at $R \sim \ell_\eta$ yields  $\varphi_2 \approx \pi / 2 - \sqrt{\beta} + \pi m$.

\begin{figure}
\vspace*{2mm} \centering
\includegraphics[width=9cm]{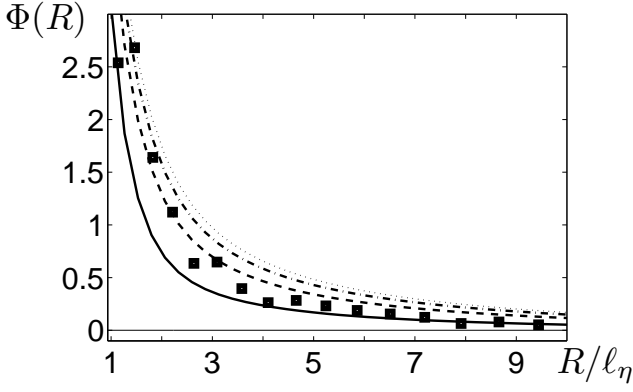}
\caption{\label{Fig4} Normalized two-point second-order correlation function $\Phi(R)$ determined from our model for $a_c=1$ and different values of the parameter $\sigma_{_{T}}$:  $\, \sigma_{_{T}}=0.5$ (solid line), $\sigma_{_{T}}=1$ (dashed line), $\sigma_{_{T}}=1.5$ (dashed-dotted line) and $\sigma_{_{T}}=2$ (dotted line). Normalized two-point second-order correlation function $\Phi(R)$ determined in our experiments (filled squares as in Fig.~\ref{Fig2}).}
\end{figure}

Now we solve Eq.~(\ref{T1}) numerically in order to determine the two-point second-order correlation function $\Phi(R)$. The normalized correlation functions of particle number density for different parameters $\sigma_{_{T}}$ are shown in Figs.~\ref{Fig2} and~\ref{Fig4}. The numerical solution for the correlation function $\Phi(R)$ is in agreement with the obtained experimental results and with the results of the asymptotic analysis.

We would like to stress that in the above analysis we have neglected the corresponding contributions which describe the pure inertial clustering (i.e., clustering without imposed mean temperature gradient), because we study the case $\alpha^2 \gg 1$, i.e., we consider the case of a small Stokes number (that in our experiments is ${\rm St} =\tau_s/\tau_\eta = 5.9 \times 10^{-2}$) and not very large Reynolds numbers (that in our experiments is ${\rm Re} = u_0 \, \ell_0 / \nu =250$). In this case contributions which determine the tangling clustering are much larger than that of the pure inertial clustering.

Indeed, the main contribution to the inertial clustering determined by the function $B({\bf R})$ in the scales $a_p < R < \ell_\eta$, can be estimated as
\begin{eqnarray}
B_{\rm inertial} &=& {20 \, \sigma_u \over (1 + \sigma_u)} \, {\rm Re}^{1/2} \sim 20 \, \sigma_u \, {\rm Re}^{1/2}
\nonumber\\
&\sim& 20 \, {u_0^4 \over c_s^4} \, {\alpha^2 \over \ln^2{\rm Re}} \, {\rm Re}^{3/2}\;,
 \label{AD11}
\end{eqnarray}
[see Eq.~(42) in \cite{EKR02}], where $c_s$ is the sound speed, and we have taken into account that $\sigma_u \sim {\rm St}^2 \ll 1$. On the other hand, the main contribution to the tangling clustering in the scales $a_p < R < \ell_\eta$ is estimated as
\begin{eqnarray}
B_{\rm tangling} \sim 30 \, {\ell_0^2 \, (\nabla_z T)^2 \over T^2} \, {\alpha^2 \over \ln^2{\rm Re}} \,\, {\rm Re} \;,
 \label{AD12}
\end{eqnarray}
where we used Eqs.~(\ref{IB3}) and~(\ref{IT3}).
Therefore, the ratio $\beta_\ast \equiv B_{\rm inertial} / B_{\rm tangling}$ is of the order of
\begin{eqnarray}
\beta_\ast \equiv {B_{\rm inertial} \over B_{\rm tangling}} \sim \, {\rm Re}^{1/2} \, {u_0^4 \over c_s^4} \, \biggl[{\ell_0 \, (\nabla_z T) \over T} \biggr]^{-2} \; .
 \label{AD14}
\end{eqnarray}
For the parameters pertinent to our experiments, the ratio $\beta_\ast \sim 5 \times 10^{-9}$ is negligibly small. This is the reason why we have neglected the inertial clustering effect in our study.
Note that for atmospheric turbulence with a strong temperature inversion (with a mean temperature gradient of the order of 1 K per 100 m) the parameter $\beta_\ast \sim 10^{-2}$, while for atmospheric turbulence with a mean temperature gradient of the order of 1 K per 1000 m, the parameter $\beta_\ast \sim 1$. In these estimates we take into account that the characteristic parameters of the
atmospheric turbulent boundary layer are: integral (maximum) scale of turbulent flow $\ell_0 \sim 10 - 100$ m; the turbulent velocity in the integral scale $u_0 \sim 0.3 - 1 $ m/s; Reynolds number ${\rm Re} \sim 10^6 - 10^7 $ (see, e.g., \cite{CSA80,BLA97}).

The small yet finite clustering effect (see unfilled circles in Fig.~\ref{Fig3}) which has been observed in our experiments in the absence of the mean temperature gradient can be explained by sedimentation of particles in the gravity field. In this case the steady state solution of the equation for the mean particle number density reads:
\begin{eqnarray}
{\bf W}_g \, N - D_{_{T}} \, \bec{\nabla} N =0
\;,
 \label{AD15}
\end{eqnarray}
where ${\bf W}_g$ is the terminal fall velocity of particles. This equation yields the following mean particle number density profile for a homogeneous turbulence:
\begin{eqnarray}
N(z) = N_0 \, \exp \Big[-{z \over L_N} \Big] \;,
 \label{AD16}
\end{eqnarray}
with the characteristic scale of the mean particle number density, $L_N = D_{_{T}} / W_g \sim 10$~cm for $10 \, \mu$m particles. Tangling of a gradient of the mean particle number density by velocity fluctuations causes particle clustering due to a combined effect of gravitational-tangling clustering. However, in the turbulence with the imposed mean temperature gradient the effect of tangling clustering is considerably stronger (see filled squares in Fig.~\ref{Fig3}).

Indeed, the dimensionless equation for the two-point second-order correlation function $\Phi(R)$ for the gravitational-tangling clustering reads:
\begin{eqnarray}
{1 \over M(R)} \Big[\Phi'' + 2 \, \Big({1 \over R} + \chi(R) \Big) \, \Phi' \Big] = - I_g(R) \;,
\label{GT1}
\end{eqnarray}
where the functions $M(R)$ and $\chi(R)$ are determined by Eqs.~(\ref{L1}) and~(\ref{L2}), and
\begin{eqnarray}
I_g(R) &=& I_0 \,  \Big[1 - {3 \over 2 M(R)} - {R \, \chi(R) \over M(R)}\Big] \;,
\label{GT2}\\
I_0 &=& {\ell_0^2 \over 4 \, L_N^2} = {9 W_g^2 \over 4 \, u_0^2} \; .
\label{GT3}
\end{eqnarray}
The solution of Eq.~(\ref{GT1}) in different ranges of scales is as follows:
\begin{eqnarray}
\Phi(R) = {I_0 \over 2 \beta_M \, {\rm Re}^{1/2}} \,  \Big[ \ln {\ell_0 \over R} - {a_p \over R} \Big] \;,
\label{GT4}
\end{eqnarray}
for the viscous range of scales $a_p \leq R < {\rm Re}^{-3/4}$,
\begin{eqnarray}
\Phi(R) = B_1 \, R^{-\kappa} - {9 I_0 \over 2 (3\kappa+2)} \, R^{2/3} \;,
\quad \kappa= {7 - \sigma \over 3 + 7\sigma} \;,
\label{GT5}
\end{eqnarray}
for the inertial range of scales ${\rm Re}^{-3/4} \leq R \leq 1$;
and $\Phi(R) = - B_2 / R$ for $R > 1$, where $B_1 \simeq 3 I_0 \, \ln  {\rm Re} / 4 {\rm Re}^{(3\kappa+2)/4} $ and $B_2 \simeq 9 I_0 / 2 (3\kappa+2)$. The correlation function  for the gravitational-tangling clustering at very small scales is $\Phi(a_n) \sim (I_0 / \sqrt{\rm Re}) \,  \ln (\ell_0 / a_p)$, that in our experiments is of the order of $\sim 0.1$. The latter value is much smaller than that for the tangling clustering in the turbulence with the imposed mean temperature gradient.

\section{Discussion and Conclusions}

In the present experimental and theoretical study we have found a new type of particle clustering, tangling clustering of inertial particles, that occurs in a stably stratified turbulence with imposed mean vertical temperature gradient. Fluctuations of the particle number density are generated by tangling of the large-scale gradient, $\bec\nabla N$, by velocity fluctuations. The gradient of the mean particle number density is formed in the stratified turbulence
with imposed mean vertical temperature gradient due to the phenomenon of turbulent thermal diffusion. The tangling clustering of inertial particles is also enhanced by the additional tangling of the mean temperature gradient by velocity fluctuations, that results in generation of the temperature fluctuations. Therefore, the heat flux due to the imposed mean temperature gradient, plays a twofold role, i.e., it causes formation of (i) the gradient of the mean particle number density and (ii) a non-zero two-point correlation function of the divergence of the particle velocity field, $B({\bf R}) = \langle \tau b({\bf x}) \, b({\bf y}) \rangle$, where $b =$ div $\, {\bf v}$. The latter is the main source of the particle clustering in small scales.

There are two contributions to the correlation function $B({\bf R})$. The first contribution is caused by particle inertia, so that the particles inside the turbulent eddies are carried out to the boundary regions between the eddies by the inertial forces. This is the main mechanism of the inertial clustering in isothermal turbulence. The second contribution to the correlation function $B({\bf R})$ is due to the temperature fluctuations produced by tangling of the mean temperature gradient by velocity fluctuations. The latter results in the tangling clustering of inertial particles. It must be emphasized that the heat flux and particle inertia are two important ingredients which cause the tangling clustering.

We have shown that in the laboratory stratified turbulence the tangling clustering is much more stronger than both, the inertial clustering and the gravitational-tangling clustering occurring in isothermal turbulence. In particular, in our experiments the correlation function for the particle clustering in isothermal turbulence is much smaller than that for the tangling clustering in non-isothermal turbulence. In the stably stratified turbulence with imposed mean temperature gradient, we have found particle clusters with the size that is of the order of several Kolmogorov length scales.

The clustering described in our study is found for inertial particles with small Stokes numbers and with material density that is much larger than the fluid density. The Reynolds numbers, based on turbulent length scale and rms velocity, ${\rm Re} =250$, in our experiments is smaller than that in experiments in isothermal turbulence described in \cite{SA08,WH05,SSA08}. Probably these are the reasons for weak inertial clustering which has been observed in our experiments. It must be emphasized that inertial and tangling clusterings in our study are understood in the statistical sense, i.e., we employ an ensemble averaging over many images with the instantaneous particle distributions rather than analyze a single instantaneous image.

In the present study we have developed a theory of the tangling clustering of inertial particles that is based on the analysis of the
two-point second-order correlation function of the particle number density. The theoretical predictions are in a good agreement with the obtained experimental results.

\medskip

\begin{acknowledgements}
We are indebted to Lance Collins and Victor L'vov for illuminating discussions. We thank Alexander Krein for his assistance in construction of the experimental set-up and Dmitry Lihovetski for his assistance in processing the experimental data. This research was supported in part by the Israel Science Foundation governed by the Israeli Academy of Sciences (Grant 259/07).
\end{acknowledgements}

\end{document}